\documentclass[aip,cha,numerical,preprint,a4paper]{revtex4-1}

\usepackage[dvipdfmx]{graphicx}%
\usepackage[dvipdfmx]{color}%
\usepackage{dcolumn}%
\usepackage{bm}%
\draft % marks overfull lines with a black rule on the right

\begin{document}

%\preprint{}

\title{Supertransmission channel for an intrinsic localized mode in a 1D nonlinear physical lattice} %Title of paper



\author{M. Sato}
\email[]{msato153@staff.kanazawa-u.ac.jp}
%\thanks{}
\author{T. Nakaguchi}
\author{T. Ishikawa}
\author{S. Shige}
\author{Y. Soga}
\affiliation{Graduate School of Natural Science and Technology, Kanazawa University\\Kanazawa, Ishikawa 920-1192, Japan}
\author{Y. Doi}
\affiliation{Graduate School of Engineering, Oosaka University\\
Suita, Oosaka 565-0871, Japan}
\author{A. J. Sievers}
\affiliation{Laboratory of Atomic and Solid State Physics, Cornell University\\
Ithaca, NY 14853-2501, USA}

\date{\today}

\begin{abstract}
It is well known that a moving intrinsic localized mode (ILM) in a nonlinear physical lattice looses energy because of the resonance between it and the underlying small amplitude plane wave spectrum. By exploring the Fourier transform (FT) properties of the nonlinear force of a running ILM in a driven and damped 1D nonlinear lattice, as described by a 2-D wavenumber and frequency map, we quantify the magnitude of the resonance where the small amplitude normal mode dispersion curve and the FT amplitude components of the ILM intersect. We show that for a traveling ILM characterized by a specific frequency and wavenumber, either inside or outside the plane wave spectrum, and for situations where both onsite and intersite nonlinearity occur, either of the hard or soft type, the strength of this resonance depends on the specific mix of the two nonlinearities. Examples are presented demonstrating that by engineering this mix the resonance can be greatly reduced. The end result is a supertransmission channel for either a driven or undriven ILM in a nonintegrable, nonlinear yet physical lattice.
\end{abstract}

\pacs{ 05.45.-a, 63.20.Pw, 05.45.Yv}% insert suggested PACS numbers in braces on next line

\maketitle %\maketitle must follow title, authors, abstract and \pacs

\begin{quotation}%==================================
Treated separately both nonlinearity and lattice discreteness have applications in many branches of physics. Domain walls, kinks and solitons are examples of important features that can appear in continuous media when the nonlinearity is large. For discrete lattices where the nonlinear excitation size can be comparable to the lattice constant only a few cases, starting with the Toda lattice model, were constructed to be integrable and, hence, to make contact with the earlier soliton dynamics. The possibility that large amplitude, localized vibrational excitations can exist in discrete nonlinear physical lattices with intersite forces was discovered nearly thirty years ago. The early work showed that there were two fundamental symmetries of such excitations: an odd symmetry for the excitation centered on a particular lattice site and an even symmetry when it was centered between two such lattice sites. Since these two symmetries have different energies the mobility of such a localized excitation is inhibited by this energy difference. In the intervening years the search for a completely mobile localized excitation in a discrete physical lattice has resulted in a shift in emphasis from trying to minimize the energy difference between the two vibrational symmetries as the excitation passes through the lattice to minimizing instead the resonance interaction that occurs between the localized excitation and the small amplitude plane wave modes. So far there has been limited success using very specific models. In this article we present a technique for minimizing such a resonance interaction for nonintegrable physical lattices. It involves generating and examining a 2-D map of the nonlinear force components in wavenumber and frequency space. We show that the Fourier transform components of different kinds of nonlinear forces give rise to real spectra, and when there is more than one type present they can be designed to cancel against each other in the region of the plane wave dispersion curve. The end result is a supertransmission channel for a localized vibrational excitation in a discrete physical lattice.
\end{quotation}

\section{Introduction}%==================================
The combination of nonlinearity and discreteness makes possible localized vibrational modes in homogeneous lattices. Such an excitation, called an intrinsic localized mode (ILM) \cite{1}, or discrete breather(DB)\cite{2,3}, can exist in a lattice of any dimension. It was shown early on that for an ILM frequency close to the plane wave normal mode spectrum mobility was possible but when far removed from the band it was stationary stable and pinned at a lattice point\cite{4}. A Peierls-Nabarro (PN) potential barrier is often introduced to characterize the fact that an ILM located halfway between two lattice sites has a different energy than one centered on a lattice site. Starting with the work on kink dynamics\cite{5} in a discrete sine-Gordon system, effort has gone into trying to remove the PN barrier for localized excitations in various discrete 1-D lattices both for practical \cite{6,7,8} and theoretical applications\cite{9,10,11,12}. For a zero velocity excitation the PN barrier has been successfully eliminated with specific parameters for 1-D discrete equations of the sine-Gordon\cite{5}, Klein-Gordon\cite{9,10,11} and nonlinear Schr\"{o}dinger \cite{13,14,15} forms. Slowly running ILMs have been shown to exist, in principle, for nonlinear intersite interactions, and for Klein-Gordon and nonlinear Schr\"{o}dinger systems with specific potentials\cite{10,16}. This has been followed by a broad effort showing that radiationless traveling localized modes can be produced with nonlinear Schr\"{o}dinger equation forms \cite{17,18,19,20,21,22,23} using specific conditions, such as stability inversion\cite{21}, saturable nonlinearity\cite{19,20,22}, and a gain-loss mechanism\cite{18}.

Another direction demonstrating localized excitation mobility makes use of the concept of moving embedded solitons (ES), where the soliton internal frequency is inside a spectrum of linear modes. Although originally identified for continuous nonlinear optical model systems that support analytical solutions\cite{17, 24,25,26}, there has even been a family of stable ES identified where analytical solutions have not been found.\cite{27} More recently localized excitations have been shown to exist in some discrete nonlinear models where they are referred to as embedded lattice solitons (ELS).\cite{17,28} The general result is that for either an ES or an ELS to exist there can be no resonance between it and the small amplitude linear waves that are also present. Although much less is known about ELS a specific example is the demonstration of a two parameter family of exact ELS for a discrete version of a complex modified Korteweg-de Vries equation that includes next-nearest-neighbor coupling.\cite{28} 

The recent focus on the resonance interaction in discrete systems between the moving nonlinear localized excitation and the plane wave spectrum as the key element\cite{SJ,GF}, rather than the PN barrier itself, has lead us to address the question how best to identify the resonance for strongly localized ILMs in discrete nonlinear physical systems where analytic solutions are unknown. We have found a practical procedure whereby the resonance strength can be readily identified and then, depending on the relative contributions of the different nonlinearities, eliminated. This is demonstrated for an ILM traveling in a driven and damped 1-D Newtonian lattice that contains both onsite and nearest neighbor (NN) intersite nonlinear forces. The approach is to focus attention on the Fourier transform (FT) of the nonlinear force presented in a 2-D $(k, \omega )$  map. Although the driven-damped condition no longer characterizes a mathematically ideal lattice, this complication is compensated by precise measurements obtained for the traveling ILM, including its shape. Using these measured shapes simulations have been carried out for traveling ILMs under the same parameter conditions but without the driver or damper. The end result for both driven and undriven lattices is the discovery of simple specific conditions for a supertransmission channel for traveling ILMs. Such discrete equations describe a variety of physical systems ranging from electrical transmission lines\cite{29}, to micromechanical arrays\cite{30,31}, to 1-D antiferromagnets\cite{32}, to nonlinear photonic crystal waveguides\cite{33}, and to other systems treated within a tight binding formulation\cite{34}. 

This paper is organized as follows. In the next section we develop the expected features of the 2-D  $(k, \omega )$  map, firstly for a simple linear traveling pulse in the continuous limit and then for two well known discrete solitons examples, namely, the Toda lattice and the Ablowitz-Ladik model. Section~\ref{chap:body} begins with mobile ILMs in the driven and damped 1-D physical lattice that contains both onsite and nearest neighbor (NN) intersite nonlinear forces. By varying the relative strengths of these two contributions the Klein-Gordon onsite nonlinear force model, the NN intersite nonlinear model and the supertransmission model for both hard and soft nonlinearity can be readily studied. A summary and conclusions are presented in Section~\ref{chap:summary}. An appendix follows.

\section{Fourier-transform map of the nonlinear force for two kinds of lattice solitons}\label{chap:2}
\subsection{Pulse in linear continuous medium}
For a display of the 2D FT map of interest consider the following 1D pulse traveling in the continuum limit
\begin{eqnarray}
\tilde \psi \left( {x,t} \right) = \tilde A\left( {x - vt} \right)\exp \left[ {i\left( {k_c x - \omega _c t} \right)} \right]
\label{eq:1}
\end{eqnarray}
\noindent where $\omega _c$  and $k_c$  are the frequency and wavenumber of a carrier wave, and $\tilde A\left( x \right)$ is the envelope. The Fourier transform in space and time of this excitation is 
\begin{eqnarray}
\tilde \psi \left( {k,\omega } \right) = 4\pi ^2 \tilde A\left( {k - k_c } \right)\delta \left( {\omega  - v\left( {k - k_c } \right) - \omega _c } \right) .
\label{eq:2}
\end{eqnarray}
\noindent Thus in $(k, \omega)$  space the amplitude components of the traveling ILM,  $\left| {\tilde \psi \left( {k,w} \right)} \right|$, are along a line $\omega  = v\left( {k - k_c } \right) + \omega _c $ centered at $k_c$  with slope  $v$.\cite{35}

\begin{figure}%-----------------fig. 1----------------------
\includegraphics{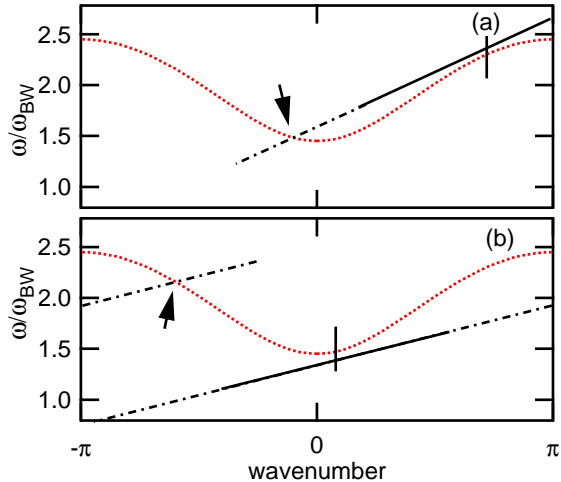}%
\caption{\label{fig:1}(a) Traveling ILM in a lattice with hard nonlinearity represented by the slanting solid lines in $(k, \omega)$  space. Dotted linear dispersion curve. Vertical line identifies the carrier wavenumber $k_c$  and its intersection with the solid line the carrier frequency $\omega _c$  of the ILM. When the localized mode amplitude becomes large and sharp in real space, the FT components spread in $k$-space (dashed line), and some interact with normal modes at the bottom of the linear dispersion curve (arrow). (b) Traveling ILM in a soft nonlinear lattice. The vertical line again identifies  $k_c$ and  $\omega _c$. Some FT components interact with normal modes near the top of the dispersion curve (arrow).}
\end{figure}

For a discrete lattice, as displayed in Fig.~\ref{fig:1}, a localized moving $\left| {\tilde \psi \left( {k,w} \right)} \right|$  in the $(k, \omega)$  space representation becomes a line centered at the carrier wavenumber $k_c$  (identified by the vertical bar) and at the carrier frequency  $\omega _c$, (the intersection of the bar and the line). Two different types of mobile ILMs are shown. Superimposed in this figure is the dispersion curve for a small amplitude Klein-Gordon (KG) lattice, represented by the dotted curve. For a traveling local mode that is sharply peaked, the FT components of  $\left| {\tilde \psi \left( {k,w} \right)} \right|$  in the $(k, \omega)$ range over a large extent in $k$-space and intersect the linear dispersion curve where the dashed and dots cross. We shall show that the nonlinear force component at this crossing accounts for the resonance between the plane wave modes and the mobile localized excitation. 

\subsection{Toda lattice soliton}%------------Toda------------------
\begin{figure}%-----------------fig. 2----------------------
\includegraphics{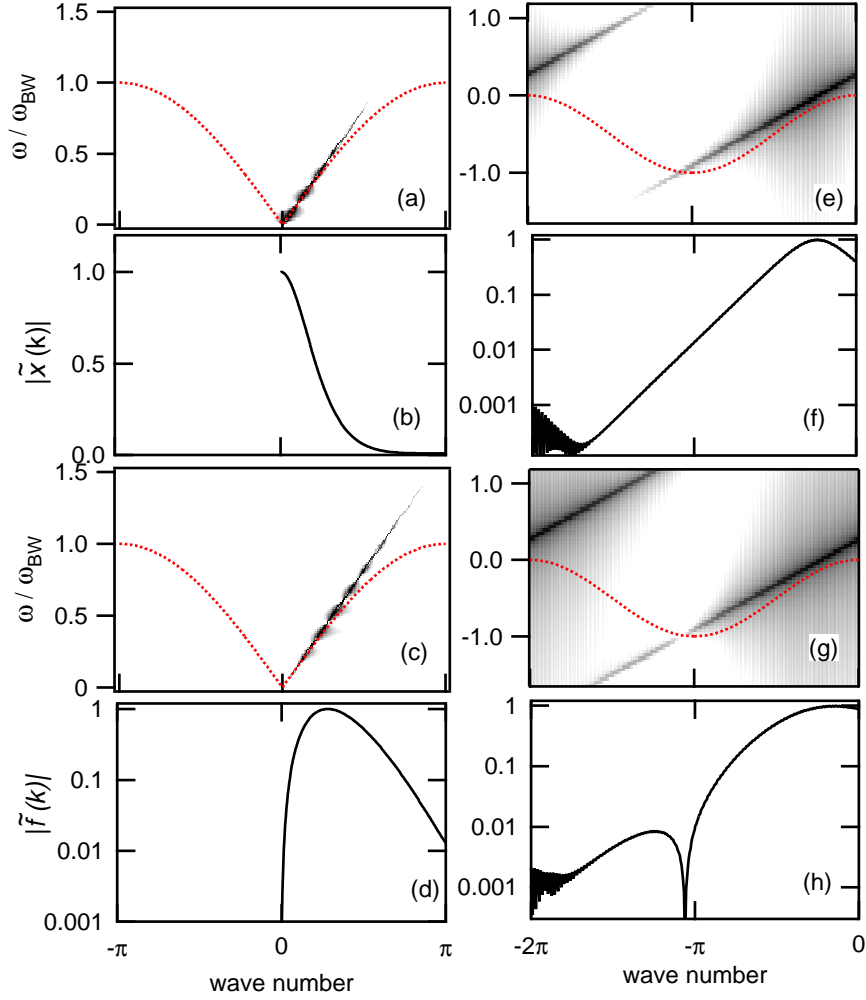}%
\caption{\label{fig:2} 2D FT analysis for Toda lattice soliton solution (a)-(d) and AL soliton solution (e)-(h). Dotted linear dispersion curves. First row panels (a),(e) show 2D-FT map of solutions $\left| {\tilde x\left( {k,\omega } \right)} \right|$. Darker gray band, larger FT amplitude. Pulsing structure in panel (a) is spurious due to finite resolution of FT. 2nd row panels (b),(f) show FT amplitude of solutions along the tangent lines $\left| {\tilde x\left( k \right)} \right|$. These amplitudes are normalized to the maxima. Panel (f), log scale. 3rd row panels (c),(g) present 2D-FT of each nonlinear force $\left| {\tilde f\left( {k,\omega } \right)} \right|$. Bottom row panels (d),(h) display FT amplitude (log scale) of the nonlinear force along the tangent line $\left| {\tilde f\left( k \right)} \right|$.
}
\end{figure}

For this special integrable lattice mobile localized excitations, first called ``lattice solitons'' by Toda\cite{36,37}, have been well studied and do not interact with the small amplitude plane wave spectrum. The purpose here is to demonstrate in the $(k, \omega)$  space representation that the FT of the nonlinear force of such an excitation does not intersect the plane wave dispersion curve. The equations of motion are characterized by
\begin{eqnarray}
\ddot x_n  = a\left\{ {2\exp \left[ { - bx_n } \right] - \exp \left[ { - bx_{n - 1} } \right] - \exp \left[ { - bx_{n + 1} } \right]} \right\},
\label{eq:3}
\end{eqnarray}
\noindent where $n$ is the lattice site,  $x_n$, the displacement, and $a$ and $b$ are nonlinear potential parameters. The linear dispersion curve $\omega  = 2\sqrt {ab} \sin \left| {k/2} \right| = \omega _{BW} \sin \left| {k/2} \right|$ is
represented by the dotted curve in Fig.~\ref{fig:2}(a), where  $\omega _{BW}  = 2\sqrt {ab} $ is the bandwidth. The single soliton solution is
\begin{eqnarray}
x_n  = \frac{1}{b}\log \frac{{\cosh \left[ {2\left( {\mu n - \omega t} \right)} \right] + \cosh 2\mu }}{{\cosh \left[ {2\left( {\mu n - \omega t} \right)} \right] + 1}}
\label{eq:4}
\end{eqnarray}
\noindent where $\omega  = \sqrt {ab} \sinh \mu $. 

To quantify the nonlinear force for the Toda equation, we linearize the force term of Eq.~(\ref{eq:3}), namely, $- ab\left\{ {2r_n  - r_{n - 1}  - r_{n + 1} } \right\}$ and then subtract it from Eq.~(\ref{eq:3}). The result is
\begin{eqnarray}
\lefteqn{ \ddot x_n  + ab\left\{ {2x_n  - x_{n - 1}  - x_{n + 1} } \right\} }\nonumber \\ 
&&  = a\left\{ {2\exp \left[ { - bx_n } \right] - \exp \left[ { - bx_{n - 1} } \right] - \exp \left[ { - bx_{n + 1} } \right]} \right\} + ab\left\{ {2x_n  - x_{n - 1}  - x_{n + 1} } \right\} 
\label{eq:5}
\end{eqnarray}
\noindent where the right hand side now describes the nonlinear force. The next step is to carry out the FT on the time dependent displacement $x_q \left( {t_p } \right)$  of the moving lattice soliton,
\begin{eqnarray}
\tilde x\left( {k_m ,\omega _n } \right) = \sum\limits_{p = 0}^{t_p  = T} {} \sum\limits_{q = 0}^{N - 1} {x_q \left( {t_p } \right)} \exp \left[ { - i\left( {k_m x_q  - \omega _n t_p } \right)} \right],
\label{eq:6}
\end{eqnarray}
\noindent where the displacement (real quantity) is first transformed over the time axis into a one-side $(0 \sim \omega _{max})$ complex spectrum. Then, the FT of the resulting complex values along the space axis gives the two sided $(k=- \pi \sim \pi )$ spectrum. The result is the  $\left| {\tilde x\left( {k,\omega } \right)} \right|$, lattice soliton amplitude map. To obtain the corresponding spectrum for the nonlinear force we insert the same numerical data,  $x_q(t_p)$, into the right hand side of Eq.~(\ref{eq:5}) to obtain the time dependent nonlinear force  $f_q(t_p)$.  Equation~(\ref{eq:6}) is then used to produce  $\left| {\tilde f\left( {k_m ,\omega _n } \right)} \right|$.

Figure~\ref{fig:2} presents the numerical illustration of these findings with the following parameters. The lattice dispersion curve is represented by the dots in Fig.~\ref{fig:2}(a) with $ab = {{\omega _{BW}^2 } \mathord{\left/
 {\vphantom {{\omega _{BW}^2 } 4}} \right. \kern-\nulldelimiterspace} 4}$  and for the soliton  $\mu =0.5$. The soliton displacement was calculated for a lattice of $N=500$  particles for a time duration $T = {\raise0.7ex\hbox{${500}$} \!\mathord{\left/
 {\vphantom {{500} {\omega _{BW} }}}\right.\kern-\nulldelimiterspace}
\!\lower0.7ex\hbox{${\omega _{BW} }$}}$  with a time step of $0.5/\omega _{BW}$ and resulting FT amplitude $\left| {\tilde x\left( {k_m ,\omega _n } \right)} \right|$  is displayed using a gray (log) scale. The maximum FT frequency $\omega _{FT\max }  = \pi \omega _{BW} $  is well beyond the maximum linear frequency  $\omega  = \omega _{BW} $. The dark straight bands represent the 2-D FT of the single soliton solution. (Darker means larger amplitude.) Because the solution is a pulse, not an envelope, the line feature starts from origin  $(k, \omega )=(0, 0)$. The line is tangent to the dispersion curve at the origin so it is the only point where the soliton line and the linear dispersion curve coincide. Panel (b) shows the FT amplitude $\left| {\tilde x\left( {k,\omega } \right)} \right|$  along that linear band. The maximum occurs at the origin. The 2-D map for the nonlinear force $\left| {\tilde f\left( {k,\omega } \right)} \right|$  is displayed in Fig.~\ref{fig:2}(c). A similar band structure to that shown in panel (a) appears for the nonlinear force but there is no longer amplitude at the origin. This feature is brought out more clearly in panel (d) where again the amplitude along the band vs wavenumber is plotted. The conclusion is that the nonlinear force spectrum is zero where the soliton spectrum and the linear dispersion curve coincide so there is no possibility of a resonance between the two kinds of excitations at finite frequency.

\subsection{Ablowitz-Ladik lattice soliton}%-----------------AL------------------
There is added value in considering the same sort of nonlinear force spectral approach for the Ablowitz-Ladik (AL) lattice, which supports a different kind of linear dispersion curve and a different kind of traveling mode solution for the single lattice soliton.\cite{38,39} The equations of motion of AL model are
\begin{eqnarray}
i\dot x_n  + K\left( {x_{n + 1}  + x_{n - 1}  - 2x_n } \right) =  - \frac{1}{2}\lambda \left| {x_n } \right|^2 \left( {x_{n + 1}  + x_{n - 1} } \right)
\label{eq:7}
\end{eqnarray}
\noindent where $K$ determines the bandwidth of the linear spectrum and  $\lambda $ is the nonlinear parameter. The dispersion curve is $\omega  = 2K\left( {\cos k - 1} \right)$ and thus $\omega <0$  for  $K>0$.  We consider  $\lambda >0$ to provide solitons. The single soliton solution is
\begin{eqnarray}
x_n \left( t \right) = \sqrt {\frac{{2K}}{\lambda }} \frac{{\sinh \mu \exp \left[ {ik_c \left( {n - vt} \right) - i\omega _c t} \right]}}{{\cosh \left[ {\mu \left( {n - vt} \right)} \right]}}
\label{eq:8}
\end{eqnarray}
\noindent where $k_c$  and  $\omega _c$ are the wave number and frequency of the carrier part of the solution, and $\mu $ and $v$  are width parameter and velocity, respectively. To analyze the solution, first we choose $k_c$  and $\omega _c$, then $\mu$ and $v$  are calculated as follows:
\begin{eqnarray}
\mu  = \cosh ^{ - 1} \left[ {\frac{1}{{\cos k_c }}\left( {\frac{{\omega _c }}{{2K}} + 1} \right)} \right],
\label{eq:9}
\end{eqnarray}
\begin{eqnarray}
v = 2K\sin k_c \frac{{\sinh \mu }}{\mu }.
\label{eq:10}
\end{eqnarray}
Inserting these into Eq.~(\ref{eq:8}), gives $x_n(t)$  over the fixed time interval with the appropriate time step as was just discussed for the Toda lattice soliton.

To illustrate the soliton and nonlinear force spectral results a specific set of parameters is now chosen; namely, $k_c  =  - \frac{{12}}{{50}}\pi $ ,  $\omega _c=-0.0278\omega _{BW}$, where the bandwidth in this case is $\omega _{BW}=4K$. The dotted curve in Fig.~\ref{fig:2}(e) identifies the dispersion curve. The wavenumber region $k=-2\pi \sim 0$  is shown instead of $k=-\pi \sim \pi$ , in order to display the entire structure of the traveling soliton. (In this case the frequency spreads both positively and negatively, because of the complex displacement in Eq.~(\ref{eq:8}).) The FT of the traveling mode is again mapped using a gray scale, where darker means larger amplitude. The dark narrow band tangent to the dispersion curve in Fig.~\ref{fig:2}(e) identifies the soliton spectrum. Figure~\ref{fig:2}(f) shows the FT amplitude along that tangent line. The peak amplitude is at the carrier wave-number and it decreases smoothly away from that location. The right hand side of Eq.~(\ref{eq:7}) is the nonlinear force. When the same procedure is carried out here as was described for the Toda lattice one finds the nonlinear force spectrum given in Fig.~\ref{fig:2}(g). Now the dark band fades away where it crosses the linear dispersion curve. Perhaps more convincing is the plot of the nonlinear force amplitude along this band shown in panel (h) where the sharp dent implies zero nonlinear amplitude at the crossing point.

\section{Nonlinear physical systems}\label{chap:body}%----------other, main part-----------
\subsection{Equations of motion}
The particular physical lattice system we have in mind is the 1-D micromechanical array, which has both onsite and intersite forces. It has been demonstrated experimentally that both components can have nonlinear contributions and that intrinsic localized modes (ILMs) can be readily produced\cite{30}. As mentioned in the introduction there are a number of different kinds of physical systems that obey this kind of nonlinear, nonintegrable equation so its choice is not at all restrictive. To describe the general case appropriate to such a 1-D physical system periodic boundary conditions are introduced for the nonlinear equations of motion for a mono-element lattice containing both onsite and nearest neighbor (NN) intersite force terms:
\begin{eqnarray}
&&\ddot x_i  + \frac{1}{\tau }\dot x_i  + \omega _O^2 x_i  + \omega _I^2 \left( {2x_i  - x_{i + 1}  - x_{i - 1} } \right) \nonumber  \\ 
&&\pm \Lambda \left( {1 - g} \right)x_i ^3   \pm \Lambda g\eta \left\{ {\left( {x_i  - x_{i + 1} } \right)^3  + \left( {x_i  - x_{i - 1} } \right)^3 } \right\} \nonumber  \\
&&= \alpha \cos \left( {k_c x_i  - \omega _c t} \right) .
\label{eq:11}
\end{eqnarray}
\noindent With $\Lambda=0$ the linear dispersion curve is $\omega ^2  = \omega _O^2  + 2\omega _I^2 \left( {1 - \cos k} \right)$, already described by the dotted curve in Fig.~\ref{fig:1}. In this example, the onsite resonance frequency  $\omega _O  = 1.45\omega _{BW} $, the bandwidth frequency $\omega _{BW}  = \sqrt {\omega _O^2  + 4\omega _I^2 }  - \omega _O $  and the normalized intersite frequency  $\omega _I  = 0.99\omega _{BW} $ all have similar values. The $\left(  \pm  \right)$  sign of 5th and 6th terms of Eq.~(\ref{eq:11}) describe positive (hard) and negative (soft) nonlinearities, respectively. Here $g$ is the fractional strength of the nonlinear intersite force and $(1-g)$, the onsite nonlinear strength. The factor $\eta $  is used to balance the strength of the two nonlinearities at a particular $k$-value, e.g., the intersite nonlinearity is more effective than the onsite one when the carrier wave is close to the zone boundary  $k=\pi $. This factor is determined from  $\eta  = \sum {x_n^4 } /\sum {\left( {x_n  - x_{n - 1} } \right)} ^4 $ where $x_n  = x_0 \cos k_c n$  is evaluated at the carrier wave number $k_c$. The right hand side of Eq.~(\ref{eq:11}) is for a propagating wave driver, which excites only the target carrier wave mode. Three parameters are borrowed from an actual cantilever array study\cite{30}: the damping term $\tau  = 8.75\left( {ms} \right) = 2770\omega _{BW} $, giving a quality factor  $Q=6400$; the nonlinear coefficient $\Lambda  = 6.5 \times 10^{20} \left( {s^{ - 2} m^{ - 2} } \right)$  and the driver amplitude  $\alpha  = 1000\left( {m/s^2 } \right)$.  The lattice size is $N=400$. 

Since there is no analytic solution for an ILM in this lattice we need a different procedure than was used earlier to generate $x_n(t)$ for the lattice soliton examples. This procedure is to produce a running ILM in the numerical simulations first by choosing  $k_c$ then finding the normal mode frequency  $\omega _n$ at  $k_c$. Next slowly increase the driver frequency from   $\omega _n$, which automatically increases the ILM amplitude: the larger the frequency difference from the normal mode, the larger the ILM amplitude and the narrower it is in real space. In steady state one obtains the time dependent displacement $x_n(t_p)$  for Eq.~(\ref{eq:6}). It is first transformed over the time axis into one-side $(0 \sim \omega _{max})$ complex spectra. A long time interval $T = {\raise0.7ex\hbox{${6280}$} \!\mathord{\left/
 {\vphantom {{6280} {\omega _{BW} }}}\right.\kern-\nulldelimiterspace}
\!\lower0.7ex\hbox{${\omega _{BW} }$}}$  is used for fine frequency resolution, and a small time step for a large maximum frequency $(\omega _{max}=20\omega _{BW})$. Then, the FT of the resulting complex values along space axis gives the two sided $(k=-\pi \sim \pi )$ spectrum. 

The FT displacement spectrum along the tangent line can be calculated from the envelope with a 1D-FT as indicated by Eqs.~(\ref{eq:1}, \ref{eq:2}). Similarly, the nonlinear force components on the 2D-FT map along the tangent line can be calculated from only the envelope by using a 1D-FT. The FT components of the onsite and intersite nonlinear forces represent real spectra as shown in the appendix and can cancel against one another when they have opposite sign.

\subsection{Results}
\subsubsection{Klein-Gordon lattice}%--------------KG-------------
\begin{figure}%-----------------------fig3---------------------
\includegraphics[height=11.5cm]{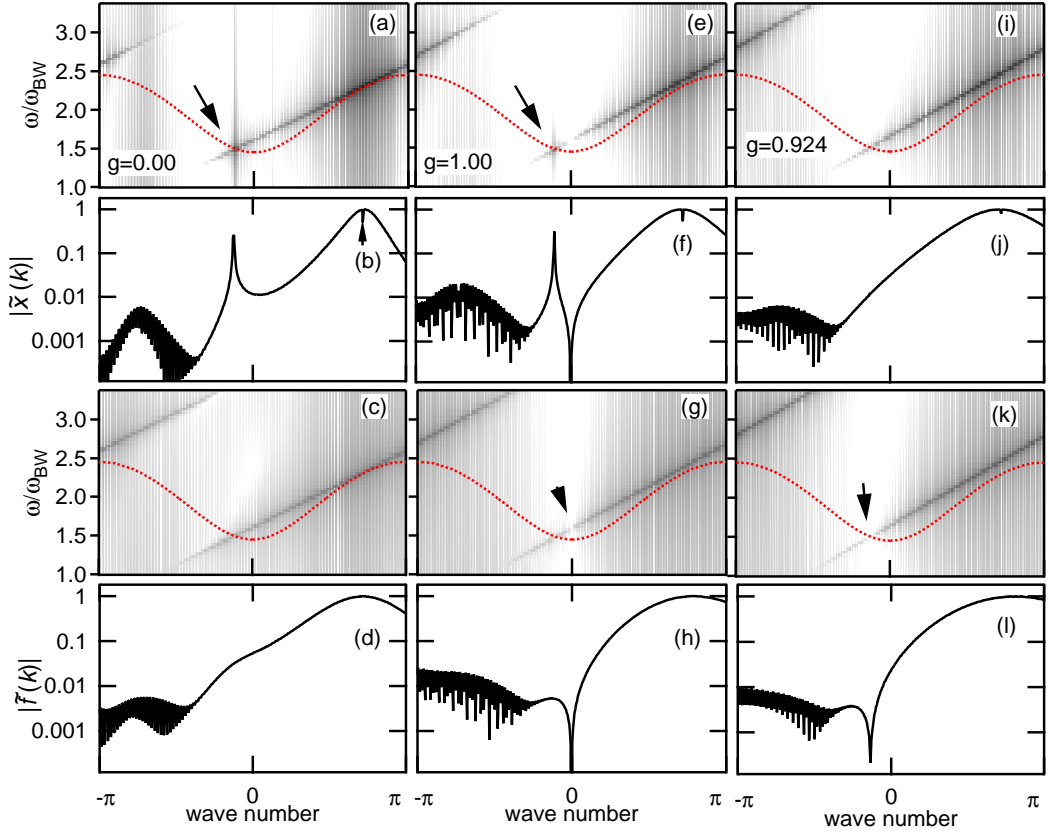}%
\caption{\label{fig:3}Two-dimensional Fourier transform (2D FT) analysis of traveling ILMs for three different models. (a)-(d) KG model; (e)-(h) NN intersite nonlinear model; and (i)-(l) a supertransmission model. Traveling modes are simulated in steady state using the driven-damped arrangement. Number of lattice sites $N=400$. (a) ILM 2D FT amplitude  $\left| {\tilde x\left( {k,\omega } \right)} \right|$ for the KG model ($g=0.00$). Darker gray identifies larger FT amplitude. Dotted trace: linear dispersion curve. Arrow locates resonance where tangent band crosses the dispersion curve. (b) 2D FT amplitude (log scale) along the center of the tangent band  $\left| {\tilde x\left( k \right)} \right|$. A sharp resonance peak is apparent. The broad peak for $k>0$ is due to the FT of the ILM envelope. The arrow centered there identifies a sharp depression due to the lattice response of the driver. The noise for $k<0$ is spurious, see text. (c) 2D FT of the nonlinear force $\left| {\tilde f\left( {k,\omega } \right)} \right|$  calculated from the displacement in (a). (d) Corresponding FT amplitude $\left| {\tilde f\left( k \right)} \right|$  (log scale) along the center of the tangent band in (c). Nonlinear force amplitude occurs at the crossing point. Frames (e) \& (i): 2D FT of the amplitude for the NN model ($g=1.00$) and mixed model ($g=0.924$). Arrow identifies the resonance in frame (e), no resonance in frame (i). Frames (f) \& (j): 2D FT amplitude (log scale) along the tangent band for $g=1.00$ and $g=0.924$. Both a resonance and antiresonance are evident in frame (f). Frames (g) \& (k): 2D FT of the nonlinear force for $g=1.00$ and $g=0.924$. The arrows show the movement of the antiresoance with the parameter change. Frames (h) \& (l): 2D FT intensity (log scale) along the center of the tangent band shown in (g) and (k). The lack of resonance intensity at the crossing point in (i) is due to the absence of the nonlinear force components at the crossing point in ($k$). 
}
\end{figure}

The ILM example presented in Fig.~\ref{fig:3}(a) shows the resulting FT amplitude $\left| {\tilde x\left( {k_m ,\omega _n } \right)} \right|$  as a function of $(k, \omega )$  for the Klein-Gordon (KG) model with onsite nonlinear coupling ($g = 0$). Here $k_c  = 144\pi /200 = 2.2619$  is one of available $k$ values for the $N=400$ lattice. As expected the FT amplitude band is centered on a line tangent to the linear dispersion curve. The arrow identifies where a resonance occurs between the FT amplitude components and normal modes of linear dispersion curve. Figure~\ref{fig:3}(b) brings out more clearly the FT amplitude along the center of the tangent band. A strong resonance is apparent. The traveling ILM looses energy by generating waves in the normal mode spectrum. The broad amplitude peak on the right hand side centered at $k_c=2.2619$  is associated with the traveling ILM. The sharp depression centered at  $k_c$ (arrow) is due to negative response of the lattice away from the ILM spatial region. Such canceling has been reported for a stationary ILM.\cite{40} (The noisy structure for  $k<0$ is produced by the large envelope signal at higher frequency, which is continued from $k=\pi $, and due to the finite FT time.) Figure~\ref{fig:3}(c) shows the 2D FT amplitude spectrum of the nonlinear force $ - \Lambda x_n ^3 $  calculated from $x_n$  in the equations of motion. The amplitude spectrum of the nonlinear force intersects the normal mode spectrum but frame (d) shows that only a monotonic dependence occurs in this region. 

\subsubsection{Nearest neighbor nonlinear lattice}%-------------NN----------
For the pure NN intersite nonlinear lattice ($g=1.0$) the 2D FT amplitude results for the ILM presented in Fig.~\ref{fig:3}(e-h) are somewhat different. Again a resonance occurs in frame (e) where the FT amplitude along the tangent band intersects the plane wave dispersion curve but nearby there is a hole in this amplitude band. The FT amplitude along the center of the tangent band displayed in frame (f) shows that both a resonance and antiresonance are evident. We now turn our attention to the properties of the antiresonance. Frame (g) presents the 2D FT amplitude spectrum of the nonlinear force $ - \Lambda \eta \left\{ {\left( {x_i  - x_{i + 1} } \right)^3  + \left( {x_i  - x_{i - 1} } \right)^3 } \right\}$ calculated from  $x_n$ in the equations of motion. The arrow identifies the location of the missing amplitude components. Again looking at the FT amplitude along the center of the tangent band, frame (h), it is evident that the strong antiresonance is a feature of the nonlinear force.

\subsubsection{Supertransmission cases}%-----------cases--------------
Qualitatively new results are obtained if both NN intersite and onsite nonlinear forces are included. Frames (i) and (j) demonstrate that by tuning the mixing ratio between the two nonlinear force terms in Eq.~(\ref{eq:11}) to $g = 0.924$ the amplitude resonance shown in frames (a)-(b) and the resonance and antiresonance shown in frames (e)-(f) are eliminated. How this occurs is illustrated in frames (k) and (l). Changing the mixing ratio of the two kinds of nonlinear force terms shifts the hole position of the nonlinear force components displayed in frame (g) to the position shown in frame (h). For this specific set of parameters in Eq.~(\ref{eq:11}) a supertransmission channel has appeared since there is no emission of plane waves produced by the propagating ILM.

Supertransmission parameters for ILMs with different sets of $k_c$ and  $\omega _c$, covering both ILMs located outside the plane wave spectrum ($1.45\omega _{BW}\sim 2.45\omega _{BW}$) as well as inside, are given in Table~\ref{table:1}. Because of $k_c$ and $\omega _c$  dependencies of the resonant crossing point, the mixing ratio $g$ must be tuned for these positive (hard) nonlinear cases  $\omega _c>\omega _n$. Examining this table shows that a smaller $g$ is required for a larger $k_c$ for this positive nonlinear case. Tuning is accomplished by adjusting the onsite term, because of the small $\eta$  (high effectiveness of intersite nonlinear term) for these  $k_c$. In all cases the mode shape in real space can be described by a carrier wave times an envelope. The last column in Table~\ref{table:1} presents the real space-time average full width at half maximum (FWHM) for each case. As expected the larger the difference in frequency between the ILM carrier and the plane wave normal mode from which it originated, the sharper the real space pattern.

%---------------------TABLE----------------------
\begin{table*}
\begin{center}
\caption{\label{table:1}Parameters for supertransmission channel.
A traveling ILM identified by $k_c$ and $\omega _c$, the carrier wave number and frequency; $\omega _n$  and $\eta$  the corresponding normal mode frequency and balancing parameter between onsite and intersite nonlinear terms.  $\omega_{BW}$ is the bandwidth and $T_{BW}=2\pi /\omega_{BW}$; $v$ and $v_g$  are traveling velocity and the (linear) group velocity at $k_c$. The last column is the time averaged full width at half maximum in real space.
}
\begin{tabular}{ccccccc}
\hline
$k_c$ & $\omega_c/\omega_n$ & $\omega_c/\omega_{BW}$ & $\eta$ & $g$ & $v\times T_{BW} (v_g\times T_{BW})$ & FWHM \\
(radians) &  &  &  &  & &(lattice constant) \\
\hline
2.13623 & 1.053 & 2.383 & 0.106 & 0.956 & 2.48(2.28) & 2.96 \\
2.2619 & 1.011 & 2.326 & 0.0932 & 0.00(KG) & 1.93(2.05) & 5.32 \\
2.2619 & 1.040 & 2.393 & 0.0932 & 1.00(NN) & 2.19(2.05) & 3.51 \\
2.2619 & 1.078 & 2.482 & 0.0932 & 0.924 & 2.30(2.05) & 2.66 \\
2.7646 & 1.140 & 2.760 & 0.0671 & 0.882 & 1.13(0.930) & 2.30 \\
0.25133* & 0.932 & 1.370 & 253 & 0.00381 & 0.939(1.03) & 4.50 \\
\hline
\multicolumn{6}{l}{\small *This is for the negative (soft) nonlinear case.}
\end{tabular}
\end{center}
\end{table*}

\subsubsection{Supertransmission without driver}%-------no driver------------------
\begin{figure}%-----------------------fig4---------------------
\includegraphics{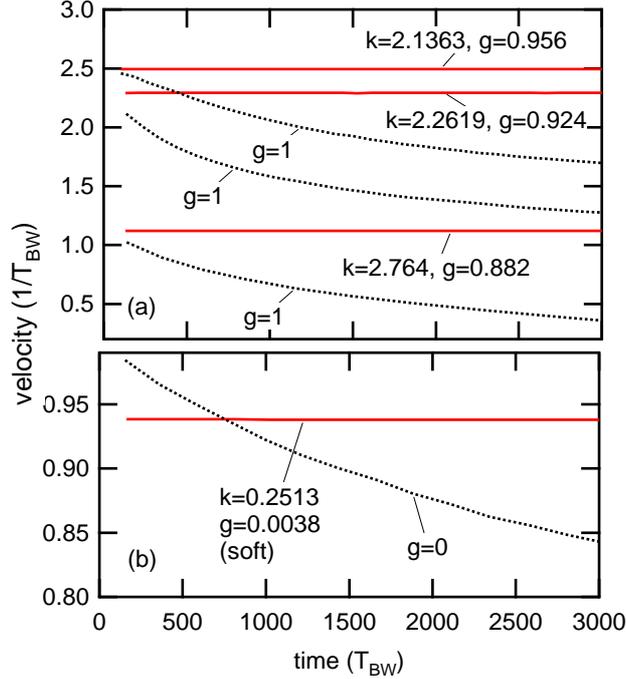}%
\caption{\label{fig:4}(a) Velocity versus time for traveling ILMs in super transmission lattice channels (solid) and corresponding regular ($g=1$) transmission lattice channels (dashed). Three cases for hard nonlinearity are shown. The traveling ILMs are simulated with the no-drive, zero damped condition. $N=3200$. Unit of time  $T_{BW}  = 2\pi /\omega _{BW} $. Each super transmission case starts from a set of displacements and velocities obtained from the corresponding driven-damped simulation. Because of the different crossing points, the tuned $g$ values are different in each case. Tuned: solid lines, the velocities are constant. Untuned: dashed lines, the velocities decrease with time because of the resonance.  (b) Case for soft nonlinearity. Tuned: solid line; untuned, dashed line. 
}
\end{figure}

To test further the specific design parameters, which produce zero nonlinear force components at the resonant crossing point and hence greatly reduced radiation loss by the moving ILM, additional simulations have been carried out for traveling ILMs with a variation of Eq.~(\ref{eq:11}), now without the driver or damper. Figure~\ref{fig:4}(a) presents ILMs velocities versus time for several $k_c$  conditions, each with two different values of the positive nonlinear onsite and intersite strengths: $g=1$(pure intersite nonlinear model) and the corresponding supertransmission tuned $g$ value. Three different $k_c$ values are shown for the positive nonlinear lattice with $N=3200$. Each $t = 0$ initial condition is obtained from the steady state driven-damped simulations with $N=400$ in Eq.~(\ref{eq:11}). The other 2800 lattice points are filled with zeros. The deceleration of the ILMs observed for each of the $g=1$  cases is due to the radiation loss related to the presence of the resonance while for tuned $g$ cases no deceleration is observed. This figure shows that by tuning the mixing ratio of the onsite and intersite nonlinearity to a specific combination for an ILM with a given $(k_c, \omega _c)$  one can greatly reduce the radiation loss. Since we obtained fairly constant velocity by launching an $N=400$ lattice ILM into an $N=3200$ lattice, these results demonstrate that this super transmission channel doesn't require background excitations, which may occur in the initial condition for the ILM obtained in the driven-damped simulations.

\begin{figure}%-----------------------fig5---------------------
\includegraphics{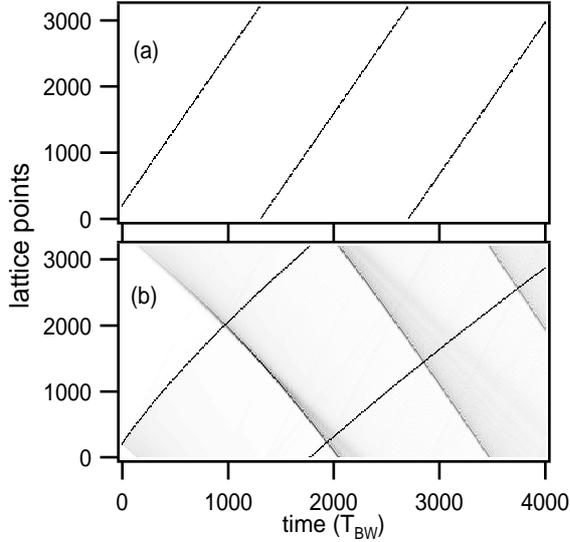}%
\caption{\label{fig:5}Real space energy density of a traveling ILM vs time. (a) the supertransmission lattice ($g=0.924$) and (b) NN lattice ($g=1$). ILM location $k_c=36\pi /50=2.2619$ in Fig.~\ref{fig:4}(a). To see the emitted waves clearly, the gray scale in each panel is magnified by 100. (a) The ILM travels at a constant speed without emitting small amplitude plane waves. (b) The plane wave emission reforms into a packet, in the opposite direction from the ILM variable speed. At $t=0$  the ILM crosses lattice site, 200.
}
\end{figure}

%-----revised part1--------
Figure~\ref{fig:5}(a, b) shows the real space energy density plot for an ILM traveling either in the supertransmission channel (a) or the corresponding lossy case (b) for a specific $k$-value. A gray scale with a magnification of 100 is used to bring out the differences between the two cases. For the supertransmission lattice, the ILM travels at a constant speed without emitting small amplitude plane waves while for the nonlinear NN lattice, plane wave emission decreases the ILM speed. Note the plane wave emission evident in (b) is in the opposite direction from that of the ILM and the accumulated waves re-form into a weak localized backward directed wave packet represented by the relatively sharp gray band.

%-----end  of revised part1--------

\subsubsection{Soft nonlinearity case}%-------------soft------------
For the negative (soft) nonlinear case ($g=0$) now describes an ILM below the bottom of the plane wave spectrum ($\omega _c<\omega _n$) for an untuned, regular KG lattice. Here tuning is accomplished by increasing the intersite term, because of the very low effectiveness of the intersite nonlinear component. Figure~\ref{fig:4}(b) shows the velocity versus time results for a soft nonlinear lattice with $k_c=0.2513$, $g=0.0038$. The parameters are given in the last row of Table~\ref{table:1}. Now the onsite nonlinearity is much more effective than the intersite nonlinearity since   $k_c$ is close to zero. In comparing the unturned case with the tuned case note that the initial velocity for the regular lattice is larger than for the tuned case because the added negative nonlinear intersite term of the tuned one reduces its ILM velocity. The much smaller deceleration observed here for the untuned case than for the other cases shown in Fig.~\ref{fig:4}(a) is partially because of a smaller ILM amplitude, and partially because of the relatively remote crossing point at $\Delta k\sim 4.7$  compared to the cases shown in Fig.~\ref{fig:4}(a) where  $\Delta k=2.3\sim 3.8$. It appears the resonance effect is more important for a fast moving ILM like that shown in Fig.~\ref{fig:1}(a) than a slow moving one like in Fig.~\ref{fig:1}(b) because for the former the intersection point of the nonlinear force Fourier component is closer to  $k_c$, the carrier point. It is also true that if the carrier frequency of the moving ILM is far enough from the band, the intersection point will occur in a much higher order Brillouin zone, producing a much weaker resonance effect so that the Peierls-Nabarro barrier would now play an important role in determining ILM mobility.

\section{Summary and Conclusions}\label{chap:summary}%==================================
%-----revised part 2-------------------
{\it Recipe for production of supertransmission channel.} Consider the positive nonlinear case. (1) With driver and damping in the equations of motion (Eq.~(\ref{eq:11})) choose  $k_c$ and find the normal mode frequency  $\omega _n$. (Initial conditions do not matter.) (2) Increase driver frequency from  $\omega _n$ to produce locked ILM. Use the minimum $\alpha $  necessary to reach the high amplitude ILM state and pass the low amplitude chaotic state quickly. In such a driven-damped condition, the nonlinear parameter $\Lambda $  can have any value  $>0$. (3) Start from $g =1$ (nonlinear NN model). (4) After driven mode reaches steady state (several 10 thousand periods) take the double FT of the time dependent displacement giving the two-sided $(k, \omega )$  spectrum. (5) Calculate the FT displacement spectrum along the tangent line. (6) Calculate the nonlinear force FT components along the tangent line. (Remember the appendix.) (7) Check spectra in intersection region for that $g$ value. (8) Change $g$ and do simulation again. (9) Repeat until zero nonlinear force is achieved. 
%------end of revised part2------------

As long as mobile ILMs can be generated and the resonance intersection occurs at one location on the $(k, \omega )$  map the range of parameter tunability to produce supertransmission extends over most of the plane wave dispersion curve shown in Fig.~\ref{fig:3}(i). For this particular dispersion curve producing supermobile ILMs in or above the top of the plane wave spectrum is straightforward but generating those below the bottom require special care. For cases near $k_c\sim \pi$  or $k_c\sim 0$  supermobility is not possible because the velocity is too small and multiple resonances occur in $k$ space. The key element for supertransmission is the presence of at least two nonlinearities, which have opposite sign where the intersection occurs in the $(k, \omega )$  map, while the sum of these nonlinearities is large enough at $k=k_c$  to produce the traveling ILM. This occurred for the physical equations considered here since the FT components of the nonlinear onsite force always have the same sign while a region of those of the NN nonlinear intersite force have opposite sign. [Note: A somewhat less physical combination we explored briefly that produced a supermobile ILM involved NN nonlinear forces together with 3rd NN nonlinear forces with a strength ratio of 1.0:0.5.] It should be noted that most physical parameters used in this paper were taken from existing microcantilever arrays but the supertransmission mixing ratios displayed here would require a larger onsite nonlinearity, hence for an experimental demonstration a redesign would be required, such as widening the head of each cantilever. 

By simulating traveling ILMs in nonintegrable 1D physical lattices with driver and damping combinations, and either hard or soft onsite and intersite nonlinearity, we have found a way to suppress the resonant interaction between the nonlinear force and the plane wave modes. Examining a running ILM, with internal frequency located either outside or inside the plane wave spectrum, in a $(k, \omega )$  map can be used to display the ILM amplitude and nonlinear force components centered on a band tangent to the normal mode dispersion curve. For both KG and NN nonlinear models we have demonstrated that the signature of a retarded ILM is characterized by a resonance in  $(k, \omega )$ space where the plane wave dispersion curve and the FT amplitude components of the ILM intersect. It was discovered that for the NN case nearby the resonance location there is an antiresonance in this FT amplitude spectrum. We have shown that for situations where both onsite and intersite nonlinearity occur the position of this hole in the nonlinear force spectrum depends on the specific mix of the two nonlinearities. By changing this mix one can move the nonlinear force hole to the amplitude intersection point, thereby removing the retardation effect for a traveling ILM characterized by a specific frequency and $k$-vector. Thus for both driven and un-driven 1-D nonlinear physical lattices with coupling of the kind shown in Eq.~(\ref{eq:11}) simple specific conditions have been found for a supertransmission channel for a traveling ILM. Similar supertransmission channel features are to be anticipated for running ILMs in 1D physical systems with more complex combinations of onsite and intersite nonlinearities. Because discrete equations of the form studied here characterize a variety of nonlinear transmission lines and other physical systems treated within a tight binding approximation such an engineered, intrinsic, low loss transmission channel is expected to be a very useful property.

\begin{acknowledgments}
M. S. was supported by JSPS-Grant-in-Aid for Scientific Research No. 25400394. A. J. S. was supported by Grant NSF-DMR-0906491 and he acknowledges the hospitality of the Department of Physics and Astronomy, University of Denver, where much of this work was completed.
\end{acknowledgments}

\appendix
\section{}
The zero nonlinear force contribution at the intersection with the linear dispersion curve is achieved by summing two force components, the nonlinear onsite and intersite terms. To see how this follows it is not necessary to consider the entire nonlinear force spectrum but only to examine the 2-D complex amplitude components along the tangent line in the $(k, \omega )$  map. Both nonlinear forces are generated from the three times multiplication of  $x_n$, which is the convolution in reciprocal space. The signal of interest is centered around  $k_c$, not $3k_c$  so the FT of the onsite term on the tangent where $\omega  = v\left( {k - k_c } \right) + \omega _c$
 is
\begin{eqnarray}
FT_{line} \left[ { - x_n ^3 } \right] =  - \tilde x\left( {k - k_c } \right)*\tilde x\left( { - k - k_c } \right)*\tilde x\left( {k - k_c } \right)
\label{eq:A1}
\end{eqnarray}
\noindent where $\tilde x\left( k \right)$ is the FT of the envelope part of $x_n$    and $*$ indicates the convolution. In our case  $\tilde x\left( k \right)$ only has a real part because of the symmetry of the envelope so that Eq.~(\ref{eq:A1}) is also real. 

A corresponding expression for the intersite term is
\begin{eqnarray}
\lefteqn {FT_{line} \left[ { - \eta \left( {x_n  - x_{n - 1} } \right)^3 } \right] } \nonumber \\
&&  =- \eta \left[ {\tilde x\left( {k - k_c } \right)\left( {1 - e^{ - ik} } \right)} \right]*\left[ {\tilde x\left( { - k - k_c } \right)\left( {1 - e^{ik} } \right)} \right]*\left[ {\tilde x\left( {k - k_c } \right)\left( {1 - e^{ - ik} } \right)} \right]
\label{eq:A2}
\end{eqnarray}
\noindent where we have made use of the fact that
\begin{eqnarray}
FT_{line} \left[ {x_n  - x_{n - 1} } \right] = \tilde x\left( {k - k_c } \right) - \tilde x\left( {k - k_c } \right)e^{ - ik}.
\label{eq:A3}
\end{eqnarray}
\noindent The end result is a complex spectrum. The other intersite term is
\begin{eqnarray}
\lefteqn {FT_{line} \left[ { - \eta \left( {x_n  - x_{n + 1} } \right)^3 } \right] } \nonumber \\
&&=  - \eta \left[ {\tilde x\left( {k - k_c } \right)\left( {1 - e^{ik} } \right)} \right]*\left[ {\tilde x\left( { - k - k_c } \right)\left( {1 - e^{ - ik} } \right)} \right]*\left[ {\tilde x\left( {k - k_c } \right)\left( {1 - e^{ik} } \right)} \right]
\label{eq:A4}
\end{eqnarray}
and when Eq.~(\ref{eq:A2}) is added to Eq.~(\ref{eq:A4}) the total intersite spectrum is real. Thus, both onsite and intersite nonlinear force spectra are real. 

%\bibliography{refsp.bib}

\begin{thebibliography}{42}%
\makeatletter
\providecommand \@ifxundefined [1]{%
 \@ifx{#1\undefined}
}%
\providecommand \@ifnum [1]{%
 \ifnum #1\expandafter \@firstoftwo
 \else \expandafter \@secondoftwo
 \fi
}%
\providecommand \@ifx [1]{%
 \ifx #1\expandafter \@firstoftwo
 \else \expandafter \@secondoftwo
 \fi
}%
\providecommand \natexlab [1]{#1}%
\providecommand \enquote  [1]{``#1''}%
\providecommand \bibnamefont  [1]{#1}%
\providecommand \bibfnamefont [1]{#1}%
\providecommand \citenamefont [1]{#1}%
\providecommand \href@noop [0]{\@secondoftwo}%
\providecommand \href [0]{\begingroup \@sanitize@url \@href}%
\providecommand \@href[1]{\@@startlink{#1}\@@href}%
\providecommand \@@href[1]{\endgroup#1\@@endlink}%
\providecommand \@sanitize@url [0]{\catcode `\\12\catcode `\$12\catcode
  `\&12\catcode `\#12\catcode `\^12\catcode `\_12\catcode `\%12\relax}%
\providecommand \@@startlink[1]{}%
\providecommand \@@endlink[0]{}%
\providecommand \url  [0]{\begingroup\@sanitize@url \@url }%
\providecommand \@url [1]{\endgroup\@href {#1}{\urlprefix }}%
\providecommand \urlprefix  [0]{URL }%
\providecommand \Eprint [0]{\href }%
\providecommand \doibase [0]{http://dx.doi.org/}%
\providecommand \selectlanguage [0]{\@gobble}%
\providecommand \bibinfo  [0]{\@secondoftwo}%
\providecommand \bibfield  [0]{\@secondoftwo}%
\providecommand \translation [1]{[#1]}%
\providecommand \BibitemOpen [0]{}%
\providecommand \bibitemStop [0]{}%
\providecommand \bibitemNoStop [0]{.\EOS\space}%
\providecommand \EOS [0]{\spacefactor3000\relax}%
\providecommand \BibitemShut  [1]{\csname bibitem#1\endcsname}%
\let\auto@bib@innerbib\@empty
%</preamble>
\bibitem [{\citenamefont {Sievers}\ and\ \citenamefont {Takeno}(1988)}]{1}%
  \BibitemOpen
  \bibfield  {author} {\bibinfo {author} {\bibfnamefont {A.~J.}\ \bibnamefont
  {Sievers}}\ and\ \bibinfo {author} {\bibfnamefont {S.}~\bibnamefont
  {Takeno}},\ }\href@noop {} {\bibfield  {journal} {\bibinfo  {journal} {Phys.
  Rev. Lett.}\ }\textbf {\bibinfo {volume} {61}},\ \bibinfo {pages} {970}
  (\bibinfo {year} {1988})}\BibitemShut {NoStop}%
\bibitem [{\citenamefont {Flach}\ and\ \citenamefont {Willis}(1998)}]{2}%
  \BibitemOpen
  \bibfield  {author} {\bibinfo {author} {\bibfnamefont {S.}~\bibnamefont
  {Flach}}\ and\ \bibinfo {author} {\bibfnamefont {C.~R.}\ \bibnamefont
  {Willis}},\ }\href@noop {} {\bibfield  {journal} {\bibinfo  {journal} {Phys.
  Rep.}\ }\textbf {\bibinfo {volume} {295}},\ \bibinfo {pages} {182} (\bibinfo
  {year} {1998})}\BibitemShut {NoStop}%
\bibitem [{\citenamefont {Flach}\ and\ \citenamefont {Gorbach}(2008)}]{3}%
  \BibitemOpen
  \bibfield  {author} {\bibinfo {author} {\bibfnamefont {S.}~\bibnamefont
  {Flach}}\ and\ \bibinfo {author} {\bibfnamefont {A.~V.}\ \bibnamefont
  {Gorbach}},\ }\href@noop {} {\bibfield  {journal} {\bibinfo  {journal} {Phys.
  Rep.}\ }\textbf {\bibinfo {volume} {467}},\ \bibinfo {pages} {1} (\bibinfo
  {year} {2008})}\BibitemShut {NoStop}%
\bibitem [{\citenamefont {Bickham}, \citenamefont {Sievers},\ and\
  \citenamefont {Takeno}(1992)}]{4}%
  \BibitemOpen
  \bibfield  {author} {\bibinfo {author} {\bibfnamefont {S.~R.}\ \bibnamefont
  {Bickham}}, \bibinfo {author} {\bibfnamefont {A.~J.}\ \bibnamefont
  {Sievers}}, \ and\ \bibinfo {author} {\bibfnamefont {S.}~\bibnamefont
  {Takeno}},\ }\href@noop {} {\bibfield  {journal} {\bibinfo  {journal} {Phys.
  Rev. B}\ }\textbf {\bibinfo {volume} {45}},\ \bibinfo {pages} {10344}
  (\bibinfo {year} {1992})}\BibitemShut {NoStop}%
\bibitem [{\citenamefont {Speight}\ and\ \citenamefont {Ward}(1994)}]{5}%
  \BibitemOpen
  \bibfield  {author} {\bibinfo {author} {\bibfnamefont {J.~M.}\ \bibnamefont
  {Speight}}\ and\ \bibinfo {author} {\bibfnamefont {R.~S.}\ \bibnamefont
  {Ward}},\ }\href@noop {} {\bibfield  {journal} {\bibinfo  {journal}
  {Nonlinearity}\ }\textbf {\bibinfo {volume} {7}},\ \bibinfo {pages} {475}
  (\bibinfo {year} {1994})}\BibitemShut {NoStop}%
\bibitem [{\citenamefont {English}\ \emph {et~al.}(2010)\citenamefont
  {English}, \citenamefont {Palmero}, \citenamefont {Sievers}, \citenamefont
  {Kevrekidis},\ and\ \citenamefont {Barnak}}]{6}%
  \BibitemOpen
  \bibfield  {author} {\bibinfo {author} {\bibfnamefont {L.~Q.}\ \bibnamefont
  {English}}, \bibinfo {author} {\bibfnamefont {F.}~\bibnamefont {Palmero}},
  \bibinfo {author} {\bibfnamefont {A.~J.}\ \bibnamefont {Sievers}}, \bibinfo
  {author} {\bibfnamefont {P.~G.}\ \bibnamefont {Kevrekidis}}, \ and\ \bibinfo
  {author} {\bibfnamefont {D.~H.}\ \bibnamefont {Barnak}},\ }\href@noop {}
  {\bibfield  {journal} {\bibinfo  {journal} {Phys. Rev. E}\ }\textbf {\bibinfo
  {volume} {81}},\ \bibinfo {pages} {046605} (\bibinfo {year}
  {2010})}\BibitemShut {NoStop}%
\bibitem [{\citenamefont {Kimura}\ and\ \citenamefont {Hikihara}(2009)}]{7}%
  \BibitemOpen
  \bibfield  {author} {\bibinfo {author} {\bibfnamefont {M.}~\bibnamefont
  {Kimura}}\ and\ \bibinfo {author} {\bibfnamefont {T.}~\bibnamefont
  {Hikihara}},\ }\href@noop {} {\bibfield  {journal} {\bibinfo  {journal}
  {Chaos}\ }\textbf {\bibinfo {volume} {19}},\ \bibinfo {pages} {013138}
  (\bibinfo {year} {2009})}\BibitemShut {NoStop}%
\bibitem [{\citenamefont {Watanabe}, \citenamefont {Hamada},\ and\
  \citenamefont {Sugimoto}(2012)}]{8}%
  \BibitemOpen
  \bibfield  {author} {\bibinfo {author} {\bibfnamefont {Y.}~\bibnamefont
  {Watanabe}}, \bibinfo {author} {\bibfnamefont {K.}~\bibnamefont {Hamada}}, \
  and\ \bibinfo {author} {\bibfnamefont {N.}~\bibnamefont {Sugimoto}},\
  }\href@noop {} {\bibfield  {journal} {\bibinfo  {journal} {J. Phys. Soc.
  Jpn.}\ }\textbf {\bibinfo {volume} {81}},\ \bibinfo {pages} {014002}
  (\bibinfo {year} {2012})}\BibitemShut {NoStop}%
\bibitem [{\citenamefont {Dmitriev}\ \emph
  {et~al.}(2006{\natexlab{a}})\citenamefont {Dmitriev}, \citenamefont
  {Kevrekidis}, \citenamefont {Yoshikawa},\ and\ \citenamefont
  {Frantzeskakis}}]{9}%
  \BibitemOpen
  \bibfield  {author} {\bibinfo {author} {\bibfnamefont {S.~V.}\ \bibnamefont
  {Dmitriev}}, \bibinfo {author} {\bibfnamefont {P.~G.}\ \bibnamefont
  {Kevrekidis}}, \bibinfo {author} {\bibfnamefont {N.}~\bibnamefont
  {Yoshikawa}}, \ and\ \bibinfo {author} {\bibfnamefont {D.~J.}\ \bibnamefont
  {Frantzeskakis}},\ }\href@noop {} {\bibfield  {journal} {\bibinfo  {journal}
  {Phys. Rev. E}\ }\textbf {\bibinfo {volume} {74}},\ \bibinfo {pages} {046609}
  (\bibinfo {year} {2006}{\natexlab{a}})}\BibitemShut {NoStop}%
\bibitem [{\citenamefont {Flach}, \citenamefont {Zolotaryuk},\ and\
  \citenamefont {Kladko}(1999)}]{10}%
  \BibitemOpen
  \bibfield  {author} {\bibinfo {author} {\bibfnamefont {S.}~\bibnamefont
  {Flach}}, \bibinfo {author} {\bibfnamefont {Y.}~\bibnamefont {Zolotaryuk}}, \
  and\ \bibinfo {author} {\bibfnamefont {K.}~\bibnamefont {Kladko}},\
  }\href@noop {} {\bibfield  {journal} {\bibinfo  {journal} {Phys. Rev. E}\
  }\textbf {\bibinfo {volume} {59}},\ \bibinfo {pages} {6105} (\bibinfo {year}
  {1999})}\BibitemShut {NoStop}%
\bibitem [{\citenamefont {Speight}(1997)}]{11}%
  \BibitemOpen
  \bibfield  {author} {\bibinfo {author} {\bibfnamefont {J.~M.}\ \bibnamefont
  {Speight}},\ }\href@noop {} {\bibfield  {journal} {\bibinfo  {journal}
  {Nonlinearity}\ }\textbf {\bibinfo {volume} {10}},\ \bibinfo {pages} {1615}
  (\bibinfo {year} {1997})}\BibitemShut {NoStop}%
\bibitem [{\citenamefont {Doi}\ and\ \citenamefont {Yoshimura}(2009)}]{12}%
  \BibitemOpen
  \bibfield  {author} {\bibinfo {author} {\bibfnamefont {Y.}~\bibnamefont
  {Doi}}\ and\ \bibinfo {author} {\bibfnamefont {K.}~\bibnamefont
  {Yoshimura}},\ }\href@noop {} {\bibfield  {journal} {\bibinfo  {journal} {J.
  Phys. Soc. Jpn.}\ }\textbf {\bibinfo {volume} {78}},\ \bibinfo {pages}
  {034401} (\bibinfo {year} {2009})}\BibitemShut {NoStop}%
\bibitem [{\citenamefont {Dmitriev}\ \emph
  {et~al.}(2006{\natexlab{b}})\citenamefont {Dmitriev}, \citenamefont
  {Kevrekidis}, \citenamefont {Sukhorukov}, \citenamefont {Yoshikawa},\ and\
  \citenamefont {Takeno}}]{13}%
  \BibitemOpen
  \bibfield  {author} {\bibinfo {author} {\bibfnamefont {S.~V.}\ \bibnamefont
  {Dmitriev}}, \bibinfo {author} {\bibfnamefont {P.~G.}\ \bibnamefont
  {Kevrekidis}}, \bibinfo {author} {\bibfnamefont {A.~A.}\ \bibnamefont
  {Sukhorukov}}, \bibinfo {author} {\bibfnamefont {N.}~\bibnamefont
  {Yoshikawa}}, \ and\ \bibinfo {author} {\bibfnamefont {S.}~\bibnamefont
  {Takeno}},\ }\href@noop {} {\bibfield  {journal} {\bibinfo  {journal} {Phys.
  Lett. A}\ }\textbf {\bibinfo {volume} {356}},\ \bibinfo {pages} {324}
  (\bibinfo {year} {2006}{\natexlab{b}})}\BibitemShut {NoStop}%
\bibitem [{\citenamefont {Had{\v z}ievski}\ \emph {et~al.}(2004)\citenamefont
  {Had{\v z}ievski}, \citenamefont {Maluckov}, \citenamefont {Stepi\'c},\ and\
  \citenamefont {Kip}}]{14}%
  \BibitemOpen
  \bibfield  {author} {\bibinfo {author} {\bibfnamefont {L.}~\bibnamefont
  {Had{\v z}ievski}}, \bibinfo {author} {\bibfnamefont {A.}~\bibnamefont
  {Maluckov}}, \bibinfo {author} {\bibfnamefont {M.}~\bibnamefont {Stepi\'c}},
  \ and\ \bibinfo {author} {\bibfnamefont {D.}~\bibnamefont {Kip}},\
  }\href@noop {} {\bibfield  {journal} {\bibinfo  {journal} {Phys. Rev. Lett.}\
  }\textbf {\bibinfo {volume} {93}},\ \bibinfo {pages} {033901} (\bibinfo
  {year} {2004})}\BibitemShut {NoStop}%
\bibitem [{\citenamefont {Pelinovsky}(2006)}]{15}%
  \BibitemOpen
  \bibfield  {author} {\bibinfo {author} {\bibfnamefont {D.~E.}\ \bibnamefont
  {Pelinovsky}},\ }\href@noop {} {\bibfield  {journal} {\bibinfo  {journal}
  {Nonlinearity}\ }\textbf {\bibinfo {volume} {19}},\ \bibinfo {pages} {2695}
  (\bibinfo {year} {2006})}\BibitemShut {NoStop}%
\bibitem [{\citenamefont {Flach}\ and\ \citenamefont {Kladko}(1999)}]{16}%
  \BibitemOpen
  \bibfield  {author} {\bibinfo {author} {\bibfnamefont {S.}~\bibnamefont
  {Flach}}\ and\ \bibinfo {author} {\bibfnamefont {K.}~\bibnamefont {Kladko}},\
  }\href@noop {} {\bibfield  {journal} {\bibinfo  {journal} {Physica D}\
  }\textbf {\bibinfo {volume} {127}},\ \bibinfo {pages} {61} (\bibinfo {year}
  {1999})}\BibitemShut {NoStop}%
\bibitem [{\citenamefont {Fujioka}, \citenamefont {Espinosa-Cer\'on},\ and\
  \citenamefont {Rodr\'iguez}(2006)}]{17}%
  \BibitemOpen
  \bibfield  {author} {\bibinfo {author} {\bibfnamefont {J.}~\bibnamefont
  {Fujioka}}, \bibinfo {author} {\bibfnamefont {A.}~\bibnamefont
  {Espinosa-Cer\'on}}, \ and\ \bibinfo {author} {\bibfnamefont {R.~F.}\
  \bibnamefont {Rodr\'iguez}},\ }\href@noop {} {\bibfield  {journal} {\bibinfo
  {journal} {Rev. Mex. Fis.}\ }\textbf {\bibinfo {volume} {52}},\ \bibinfo
  {pages} {6} (\bibinfo {year} {2006})}\BibitemShut {NoStop}%
\bibitem [{\citenamefont {Johansson}, \citenamefont {Prilepsky},\ and\
  \citenamefont {Derevyanko}(2014)}]{18}%
  \BibitemOpen
  \bibfield  {author} {\bibinfo {author} {\bibfnamefont {M.}~\bibnamefont
  {Johansson}}, \bibinfo {author} {\bibfnamefont {J.~E.}\ \bibnamefont
  {Prilepsky}}, \ and\ \bibinfo {author} {\bibfnamefont {S.~A.}\ \bibnamefont
  {Derevyanko}},\ }\href@noop {} {\bibfield  {journal} {\bibinfo  {journal}
  {Phys. Rev. E}\ }\textbf {\bibinfo {volume} {89}},\ \bibinfo {pages} {042912}
  (\bibinfo {year} {2014})}\BibitemShut {NoStop}%
\bibitem [{\citenamefont {Melvin}\ \emph {et~al.}(2006)\citenamefont {Melvin},
  \citenamefont {Champneys}, \citenamefont {Kevrekidis},\ and\ \citenamefont
  {Cuevas}}]{19}%
  \BibitemOpen
  \bibfield  {author} {\bibinfo {author} {\bibfnamefont {T.~R.~O.}\
  \bibnamefont {Melvin}}, \bibinfo {author} {\bibfnamefont {A.~R.}\
  \bibnamefont {Champneys}}, \bibinfo {author} {\bibfnamefont {P.~G.}\
  \bibnamefont {Kevrekidis}}, \ and\ \bibinfo {author} {\bibfnamefont
  {J.}~\bibnamefont {Cuevas}},\ }\href@noop {} {\bibfield  {journal} {\bibinfo
  {journal} {Phys. Rev. Lett.}\ }\textbf {\bibinfo {volume} {97}},\ \bibinfo
  {pages} {124101} (\bibinfo {year} {2006})}\BibitemShut {NoStop}%
\bibitem [{\citenamefont {Melvin}\ \emph {et~al.}(2008)\citenamefont {Melvin},
  \citenamefont {Champneys}, \citenamefont {Kevrekidis},\ and\ \citenamefont
  {Cuevas}}]{20}%
  \BibitemOpen
  \bibfield  {author} {\bibinfo {author} {\bibfnamefont {T.~R.~O.}\
  \bibnamefont {Melvin}}, \bibinfo {author} {\bibfnamefont {A.~R.}\
  \bibnamefont {Champneys}}, \bibinfo {author} {\bibfnamefont {P.~G.}\
  \bibnamefont {Kevrekidis}}, \ and\ \bibinfo {author} {\bibfnamefont
  {J.}~\bibnamefont {Cuevas}},\ }\href@noop {} {\bibfield  {journal} {\bibinfo
  {journal} {Physica D}\ }\textbf {\bibinfo {volume} {237}},\ \bibinfo {pages}
  {551} (\bibinfo {year} {2008})}\BibitemShut {NoStop}%
\bibitem [{\citenamefont {{\"O}ster}, \citenamefont {Johansson},\ and\
  \citenamefont {Eriksson}(2003)}]{21}%
  \BibitemOpen
  \bibfield  {author} {\bibinfo {author} {\bibfnamefont {M.}~\bibnamefont
  {{\"O}ster}}, \bibinfo {author} {\bibfnamefont {M.}~\bibnamefont
  {Johansson}}, \ and\ \bibinfo {author} {\bibfnamefont {A.}~\bibnamefont
  {Eriksson}},\ }\href@noop {} {\bibfield  {journal} {\bibinfo  {journal}
  {Phys. Rev. E}\ }\textbf {\bibinfo {volume} {67}},\ \bibinfo {pages} {056606}
  (\bibinfo {year} {2003})}\BibitemShut {NoStop}%
\bibitem [{\citenamefont {Oxtoby}\ and\ \citenamefont
  {Barashenkov}(2007)}]{22}%
  \BibitemOpen
  \bibfield  {author} {\bibinfo {author} {\bibfnamefont {O.~F.}\ \bibnamefont
  {Oxtoby}}\ and\ \bibinfo {author} {\bibfnamefont {I.~V.}\ \bibnamefont
  {Barashenkov}},\ }\href@noop {} {\bibfield  {journal} {\bibinfo  {journal}
  {Phys. Rev. E}\ }\textbf {\bibinfo {volume} {76}},\ \bibinfo {pages} {036603}
  (\bibinfo {year} {2007})}\BibitemShut {NoStop}%
\bibitem [{\citenamefont {Johansson}(2006)}]{23}%
  \BibitemOpen
  \bibfield  {author} {\bibinfo {author} {\bibfnamefont {M.}~\bibnamefont
  {Johansson}},\ }\href@noop {} {\bibfield  {journal} {\bibinfo  {journal}
  {Physica D}\ }\textbf {\bibinfo {volume} {216}},\ \bibinfo {pages} {62}
  (\bibinfo {year} {2006})}\BibitemShut {NoStop}%
\bibitem [{\citenamefont {Fujioka}\ and\ \citenamefont {Espinosa}(1997)}]{24}%
  \BibitemOpen
  \bibfield  {author} {\bibinfo {author} {\bibfnamefont {J.}~\bibnamefont
  {Fujioka}}\ and\ \bibinfo {author} {\bibfnamefont {A.}~\bibnamefont
  {Espinosa}},\ }\href@noop {} {\bibfield  {journal} {\bibinfo  {journal} {J.
  Phys. Soc. Jpn.}\ }\textbf {\bibinfo {volume} {66}},\ \bibinfo {pages} {2601}
  (\bibinfo {year} {1997})}\BibitemShut {NoStop}%
\bibitem [{\citenamefont {Champneys}, \citenamefont {Malomed},\ and\
  \citenamefont {Friedman}(1998)}]{25}%
  \BibitemOpen
  \bibfield  {author} {\bibinfo {author} {\bibfnamefont {A.~R.}\ \bibnamefont
  {Champneys}}, \bibinfo {author} {\bibfnamefont {B.~A.}\ \bibnamefont
  {Malomed}}, \ and\ \bibinfo {author} {\bibfnamefont {M.~J.}\ \bibnamefont
  {Friedman}},\ }\href@noop {} {\bibfield  {journal} {\bibinfo  {journal}
  {Phys. Rev. Lett.}\ }\textbf {\bibinfo {volume} {80}},\ \bibinfo {pages}
  {4169} (\bibinfo {year} {1998})}\BibitemShut {NoStop}%
\bibitem [{\citenamefont {Champneys}\ and\ \citenamefont {Malomed}(2000)}]{26}%
  \BibitemOpen
  \bibfield  {author} {\bibinfo {author} {\bibfnamefont {A.~R.}\ \bibnamefont
  {Champneys}}\ and\ \bibinfo {author} {\bibfnamefont {B.~A.}\ \bibnamefont
  {Malomed}},\ }\href@noop {} {\bibfield  {journal} {\bibinfo  {journal} {Phys.
  Rev. E}\ }\textbf {\bibinfo {volume} {61}},\ \bibinfo {pages} {886} (\bibinfo
  {year} {2000})}\BibitemShut {NoStop}%
\bibitem [{\citenamefont {Yang}(2003)}]{27}%
  \BibitemOpen
  \bibfield  {author} {\bibinfo {author} {\bibfnamefont {J.}~\bibnamefont
  {Yang}},\ }\href@noop {} {\bibfield  {journal} {\bibinfo  {journal} {Phys.
  Rev. Lett.}\ }\textbf {\bibinfo {volume} {91}},\ \bibinfo {pages} {143903}
  (\bibinfo {year} {2003})}\BibitemShut {NoStop}%
\bibitem [{\citenamefont {Malomed}\ \emph {et~al.}(2006)\citenamefont
  {Malomed}, \citenamefont {Fujioka}, \citenamefont {Espinosa-Cer\'on},
  \citenamefont {Rodr\'iguez},\ and\ \citenamefont {Gonz\'alez}}]{28}%
  \BibitemOpen
  \bibfield  {author} {\bibinfo {author} {\bibfnamefont {B.~A.}\ \bibnamefont
  {Malomed}}, \bibinfo {author} {\bibfnamefont {J.}~\bibnamefont {Fujioka}},
  \bibinfo {author} {\bibfnamefont {A.}~\bibnamefont {Espinosa-Cer\'on}},
  \bibinfo {author} {\bibfnamefont {R.~F.}\ \bibnamefont {Rodr\'iguez}}, \ and\
  \bibinfo {author} {\bibfnamefont {S.}~\bibnamefont {Gonz\'alez}},\
  }\href@noop {} {\bibfield  {journal} {\bibinfo  {journal} {Chaos}\ }\textbf
  {\bibinfo {volume} {16}},\ \bibinfo {pages} {013112} (\bibinfo {year}
  {2006})}\BibitemShut {NoStop}%
\bibitem [{\citenamefont {Sire}\ and\ \citenamefont {James}(2005)}]{SJ}%
  \BibitemOpen
  \bibfield  {author} {\bibinfo {author} {\bibfnamefont {Y.}~\bibnamefont
  {Sire}}\ and\ \bibinfo {author} {\bibfnamefont {G.}~\bibnamefont {James}},\
  }\href@noop {} {\bibfield  {journal} {\bibinfo  {journal} {Phyca D}\ }\textbf
  {\bibinfo {volume} {204}},\ \bibinfo {pages} {15} (\bibinfo {year}
  {2005})}\BibitemShut {NoStop}%
\bibitem [{\citenamefont {G{\'o}mez-Garde{\~ n}es}\ \emph
  {et~al.}(2004)\citenamefont {G{\'o}mez-Garde{\~ n}es}, \citenamefont
  {Flor\'ia}, \citenamefont {Peyrard},\ and\ \citenamefont {Bishop}}]{GF}%
  \BibitemOpen
  \bibfield  {author} {\bibinfo {author} {\bibfnamefont {J.}~\bibnamefont
  {G{\'o}mez-Garde{\~ n}es}}, \bibinfo {author} {\bibfnamefont {L.~M.}\
  \bibnamefont {Flor\'ia}}, \bibinfo {author} {\bibfnamefont {M.}~\bibnamefont
  {Peyrard}}, \ and\ \bibinfo {author} {\bibfnamefont {A.~R.}\ \bibnamefont
  {Bishop}},\ }\href@noop {} {\bibfield  {journal} {\bibinfo  {journal}
  {CHAOS}\ }\textbf {\bibinfo {volume} {14}},\ \bibinfo {pages} {1130}
  (\bibinfo {year} {2004})}\BibitemShut {NoStop}%
\bibitem [{\citenamefont {English}, \citenamefont {Thakur},\ and\ \citenamefont
  {Stearrett}(2008)}]{29}%
  \BibitemOpen
  \bibfield  {author} {\bibinfo {author} {\bibfnamefont {L.~Q.}\ \bibnamefont
  {English}}, \bibinfo {author} {\bibfnamefont {R.~B.}\ \bibnamefont {Thakur}},
  \ and\ \bibinfo {author} {\bibfnamefont {R.}~\bibnamefont {Stearrett}},\
  }\href@noop {} {\bibfield  {journal} {\bibinfo  {journal} {Phys. Rev. E}\
  }\textbf {\bibinfo {volume} {77}},\ \bibinfo {pages} {066601} (\bibinfo
  {year} {2008})}\BibitemShut {NoStop}%
\bibitem [{\citenamefont {Sato}, \citenamefont {Hubbard},\ and\ \citenamefont
  {Sievers}(2006)}]{30}%
  \BibitemOpen
  \bibfield  {author} {\bibinfo {author} {\bibfnamefont {M.}~\bibnamefont
  {Sato}}, \bibinfo {author} {\bibfnamefont {B.~E.}\ \bibnamefont {Hubbard}}, \
  and\ \bibinfo {author} {\bibfnamefont {A.~J.}\ \bibnamefont {Sievers}},\
  }\href@noop {} {\bibfield  {journal} {\bibinfo  {journal} {Rev. Mod. Phys.}\
  }\textbf {\bibinfo {volume} {78}},\ \bibinfo {pages} {137} (\bibinfo {year}
  {2006})}\BibitemShut {NoStop}%
\bibitem [{\citenamefont {Sato}\ \emph {et~al.}(2003)\citenamefont {Sato},
  \citenamefont {Hubbard}, \citenamefont {Sievers}, \citenamefont {Ilic},
  \citenamefont {Czaplewski},\ and\ \citenamefont {Craighead}}]{31}%
  \BibitemOpen
  \bibfield  {author} {\bibinfo {author} {\bibfnamefont {M.}~\bibnamefont
  {Sato}}, \bibinfo {author} {\bibfnamefont {B.~E.}\ \bibnamefont {Hubbard}},
  \bibinfo {author} {\bibfnamefont {A.~J.}\ \bibnamefont {Sievers}}, \bibinfo
  {author} {\bibfnamefont {B.}~\bibnamefont {Ilic}}, \bibinfo {author}
  {\bibfnamefont {D.~A.}\ \bibnamefont {Czaplewski}}, \ and\ \bibinfo {author}
  {\bibfnamefont {H.~G.}\ \bibnamefont {Craighead}},\ }\href@noop {} {\bibfield
   {journal} {\bibinfo  {journal} {Phys. Rev. Lett.}\ }\textbf {\bibinfo
  {volume} {90}},\ \bibinfo {pages} {044102} (\bibinfo {year}
  {2003})}\BibitemShut {NoStop}%
\bibitem [{\citenamefont {Lai}\ and\ \citenamefont {Sievers}(1999)}]{32}%
  \BibitemOpen
  \bibfield  {author} {\bibinfo {author} {\bibfnamefont {R.}~\bibnamefont
  {Lai}}\ and\ \bibinfo {author} {\bibfnamefont {A.~J.}\ \bibnamefont
  {Sievers}},\ }\href@noop {} {\bibfield  {journal} {\bibinfo  {journal} {Phys.
  Rep.}\ }\textbf {\bibinfo {volume} {314}},\ \bibinfo {pages} {147} (\bibinfo
  {year} {1999})}\BibitemShut {NoStop}%
\bibitem [{\citenamefont {McGurn}(1999)}]{33}%
  \BibitemOpen
  \bibfield  {author} {\bibinfo {author} {\bibfnamefont {A.~R.}\ \bibnamefont
  {McGurn}},\ }\href@noop {} {\bibfield  {journal} {\bibinfo  {journal} {Phys.
  Lett. A}\ }\textbf {\bibinfo {volume} {251}},\ \bibinfo {pages} {322}
  (\bibinfo {year} {1999})}\BibitemShut {NoStop}%
\bibitem [{\citenamefont {McGurn}\ and\ \citenamefont {Birkok}(2004)}]{34}%
  \BibitemOpen
  \bibfield  {author} {\bibinfo {author} {\bibfnamefont {A.~R.}\ \bibnamefont
  {McGurn}}\ and\ \bibinfo {author} {\bibfnamefont {G.}~\bibnamefont
  {Birkok}},\ }\href@noop {} {\bibfield  {journal} {\bibinfo  {journal} {Phys.
  Rev. B}\ }\textbf {\bibinfo {volume} {69}},\ \bibinfo {pages} {235105}
  (\bibinfo {year} {2004})}\BibitemShut {NoStop}%
\bibitem [{\citenamefont {Sato}\ and\ \citenamefont {Sievers}(2007)}]{35}%
  \BibitemOpen
  \bibfield  {author} {\bibinfo {author} {\bibfnamefont {M.}~\bibnamefont
  {Sato}}\ and\ \bibinfo {author} {\bibfnamefont {A.~J.}\ \bibnamefont
  {Sievers}},\ }\href@noop {} {\bibfield  {journal} {\bibinfo  {journal} {Phys.
  Rev. Lett.}\ }\textbf {\bibinfo {volume} {98}},\ \bibinfo {pages} {214101}
  (\bibinfo {year} {2007})}\BibitemShut {NoStop}%
\bibitem [{\citenamefont {Toda}(1967)}]{36}%
  \BibitemOpen
  \bibfield  {author} {\bibinfo {author} {\bibfnamefont {M.}~\bibnamefont
  {Toda}},\ }\href@noop {} {\bibfield  {journal} {\bibinfo  {journal} {J. Phys.
  Soc. Jpn.}\ }\textbf {\bibinfo {volume} {23}},\ \bibinfo {pages} {501}
  (\bibinfo {year} {1967})}\BibitemShut {NoStop}%
\bibitem [{\citenamefont {Toda}(1989)}]{37}%
  \BibitemOpen
  \bibfield  {author} {\bibinfo {author} {\bibfnamefont {M.}~\bibnamefont
  {Toda}},\ }\href@noop {} {\emph {\bibinfo {title} {Theory of Nonlinear
  Lattices}}}\ (\bibinfo  {publisher} {Springer-Verlag},\ \bibinfo {address}
  {New York},\ \bibinfo {year} {1989})\BibitemShut {NoStop}%
\bibitem [{\citenamefont {Ablowitz}\ and\ \citenamefont {Ladik}(1974)}]{38}%
  \BibitemOpen
  \bibfield  {author} {\bibinfo {author} {\bibfnamefont {M.~J.}\ \bibnamefont
  {Ablowitz}}\ and\ \bibinfo {author} {\bibfnamefont {J.~F.}\ \bibnamefont
  {Ladik}},\ }\href@noop {} {\bibfield  {journal} {\bibinfo  {journal} {J.
  Math. Phys.}\ }\textbf {\bibinfo {volume} {16}},\ \bibinfo {pages} {598}
  (\bibinfo {year} {1974})}\BibitemShut {NoStop}%
\bibitem [{\citenamefont {Ablowitz}\ and\ \citenamefont {Ladik}(1975)}]{39}%
  \BibitemOpen
  \bibfield  {author} {\bibinfo {author} {\bibfnamefont {M.~J.}\ \bibnamefont
  {Ablowitz}}\ and\ \bibinfo {author} {\bibfnamefont {J.~F.}\ \bibnamefont
  {Ladik}},\ }\href@noop {} {\bibfield  {journal} {\bibinfo  {journal} {J.
  Math. Phys.}\ }\textbf {\bibinfo {volume} {17}},\ \bibinfo {pages} {1011}
  (\bibinfo {year} {1975})}\BibitemShut {NoStop}%
\bibitem [{\citenamefont {R{\"o}ssler}\ and\ \citenamefont {Page}(1995)}]{40}%
  \BibitemOpen
  \bibfield  {author} {\bibinfo {author} {\bibfnamefont {T.}~\bibnamefont
  {R{\"o}ssler}}\ and\ \bibinfo {author} {\bibfnamefont {J.~B.}\ \bibnamefont
  {Page}},\ }\href@noop {} {\bibfield  {journal} {\bibinfo  {journal} {Phys.
  Lett. A}\ }\textbf {\bibinfo {volume} {204}},\ \bibinfo {pages} {418}
  (\bibinfo {year} {1995})}\BibitemShut {NoStop}%
\end{thebibliography}

\providecommand{\noopsort}[1]{}\providecommand{\singleletter}[1]{#1}%

\end{document}